# Comment on "Sign-Reversing Hall Effect in Atomically Thin High-Temperature $Bi_{2.1}Sr_{1.9}CaCu_{2.0}O_{8+\delta}$ Superconductors"

What is the fundamental reason for Hall Anomaly in Type II Superconductors: Independent Vortex Dynamics Effect or a Manifestation of Vortex Many-Body Effect? For a long time researchers in this field had not realized the importance of vortex many-body effects in the understanding of the so-called anomalous Hall effect. In this note I give a brief account of the situation in the light of a recent Phys Rev Lett.

Fabrication technologies have progressed tremendously during past two decades. Difficult experiments in 1990's can now been done much better in various ways. What reported by Zhao *et al.* [1] is such a beautiful example. To unbiased eyes, their data nicely supports the Hall anomaly theory based on the competition between pinning and vortex interaction, already partially validated in 1990's, corroborated well with another recent beautiful experiment [2]. Surprisingly, Zhao *et al.* were still formulated their interpretation within an independent vortex dynamical model whose theoretical foundation had been invalidated, and ignored competing vortex many-body model, hence reached their main conclusion wrong.

They motivated their narrative not adequately. They stated that "A rich theoretical lore attributes the Hall anomalies to either vortex pinning, details of the vortex core electronic spectrum, hydrodynamic effects, superconducting fluctuations, Berry phase, and charges in the vortex core" [1] to indicate that there were lots of theories. While their Ref.[11] was not on Hall anomaly, their brief summary actually does capture the mood in the community. As already discussed long ago [3], for Hall anomaly the inadequate physics has been evident in all those independent vortex dynamical models. Here Hall anomaly is another name for sign changes in Hall resistivity in type II superconductors. Instead, a Hall anomaly theory based on vortex many-body effect in competition with pinning was proposed [4,5] to explain such generic effect and a few quantitative predictions were obtained. It is essentially the same physics as what known in solid state physics: pinned vortex lattice behaves as filled topologically trivial band, which has zero Hall effect; the excitations are responsible for Hall effect as well as the resistance. It was explicitly tested against data published in Physical Review Letters [6]. The relevancy of such theory to Hall anomaly was immediately acknowledged [7] and supported by data from other laboratories [8-18]. The authors of [1] had not reported their comparison between their data and the vortex many-body Hall anomaly model.

To advance their theoretical model of independent vortex dynamics type, they concluded that "However, neither the explanation nor the consensus of the Hall behavior in the entire temperature range was achieved" [1]. This may be only partially valid. The vortex many-body theory not only explains one sign change, but also it predicts double or more sign changes when conditions are appropriate. In addition, the Arrhenius law behavior of the resistance is a natural outcome. Furthermore, it has two quantitative predictions which the authors may already have data to directly check against:
1) At low enough magnetic field the sign change may disappear. It is curious that they didn't present data for magnetic fields below 2T but above the effective $H_{c1}$;
2) The quantitative formulae for the "activation energy" appeared in the Arrhenius law was given [5], the energy to generate vortex vacancies in the vortex lattice, which again may be tested directly.

From a general symmetry and topology perspective it was reasoned that the Magnus force is insensitive nonmagnetic impurities in superconductors, and such result was regarded as one of very few results in many-body physics [19]. Direct experimental measurement of Magnus force on moving vortices validated such prediction [20]. Full vortex dynamics has been now obtained from microscopic BCS theory without the uncontrolled relaxation time approximation [21]. In order for their independent vortex dynamical model to apply, the Magnus force would have to be practically reduced to zero. This key ingredient in their theoretical model was achieved by a non-topological means of relaxation time approximation. It should be pointed out that the theoretical model relied on by authors was criticized long ago because of the invalidation in their relaxation time approximation. The underlying physics was discussed even earlier that a moving may disturb Fermi sea to feel a friction [22] with no need of relaxation time approximation, precisely the same situation for a moving vortex.

In conclusion, Zhao *et al*. had interpreted their main data against an invalidated independent vortex dynamics model and have reached their main conclusion incorrectly. Instead, Hall anomaly in type II superconductors is generally an effect of vortex many-body effect, not discussed by the authors but consistent with their published data and with two additional predictions.

Ping Ao
Shanghai Center for Systems Quantitative Life Sciences and Physics Department
Shanghai University
Shanghai, China
2019. 01.12